\documentclass[usenatbib]{basi}
\usepackage[T1]{fontenc}
\usepackage[british]{babel}
\usepackage[varg]{txfonts}
\usepackage[varg]{txfonts}
\usepackage{rotating}
\usepackage{dcolumn}
\usepackage{graphicx}
\begin{document}
\title[Parsec-scale structure of quasars]{Parsec-scale structure of 
	   quasars: dawn of the golden age?}
%
\author[R.~Ojha]
       {Roopesh~Ojha$^1$ $^2$\thanks{email: \texttt{Roopesh.Ojha@nasa.gov}}\\
       $^1$NASA/Goddard Space Flight Center,
       Code 661, Astroparticle Physics Laboratory,\\
       Greenbelt,, MD 20770, U.S.A. \\
      $^2$The Catholic University of America, Washington DC, U.S.A.}

\pubyear{2013}
\volume{41}
\pagerange{\pageref{firstpage}--\pageref{lastpage}}

\date{Received 2013 June 03; accepted 2013 June 21}

\maketitle
\label{firstpage}

\begin{abstract}
Half a century after their discovery, the study of quasars remains one of the 
most fascinating intellectual challenges in astronomy. Quasars are laboratories for 
everything from relativity to magnetohydrodynamics and are perhaps the 
best available probes for cosmology. A tremendous amount has been learned 
about quasars and yet many of the most fundamental questions about 
their physics remain open. 

Parsec-scale observations have played an indispensable role in 
building up our current understanding of quasars; virtually everything we know 
about quasars depends on such observations. However, the finest hour for parsec 
scale observations may be just beginning. This is partly due to the development of 
highly reliable VLBI networks (which is continuing) but mostly due to the 
unprecedented availability of multiepoch, simultaneous, broadband observations 
that have long been the `holy grail' for quasar researchers. 
%
%
\end{abstract}

\begin{keywords}
   galaxies: quasars: general -- galaxies: nuclei -- galaxies: jets -- radio continuum: galaxies
\end{keywords}

\section{Introduction}\label{s:intro}
Familiarity breeds, if not contempt, a certain complacency. Astronomers 
have become so used to obtaining parsec scale observations of objects located 
halfway across the Universe that we seldom pause to think about the magic involved. 
The word `magic' is used advisedly: the ability to combine information from radio telescopes 
separated by thousands of kilometers and produce the highest resolution images ever made 
by humans surely meets both the dictionary definitions of magic (e.g. ``a quality of being beautiful 
and delightful in a way that seems remote from daily life\footnote{Oxford Dictionaries.}") as well as 
the standard of Clarke's Third Law \citep[``Any sufficiently advanced technology is indistinguishable 
from magic."][]{1984Profiles}.  

Equally magical is what these observations have taught us about the nature of what are among the 
most mysterious and intriguing objects in the Universe: quasars. Quasars force us to reach for 
superlatives. They are enormous in extent, produce vast amounts of energy at all wavelengths, at 
incredible rates and from relatively small regions. They routinely accelerate significant fractions of 
a solar mass to near light speeds collimating the emission from this matter (whose exact nature remains 
an open question) in the direction of motion. The resulting `jets' are visible over great distances and often 
create the illusion of superluminal motion. Apart from their intrinsic interest, quasars are extremely useful 
probes in many astrophysical contexts such as the nature of the interstellar and intergalactic medium. 
They also appear to hold the key to understanding the evolutionary tracks followed by all galaxies and 
clusters of galaxies. 

Here we take a brief look at both these kinds of magic. We will look at what makes high quality 
parsec-scale observations possible, summarize what they have taught us about quasars and end with a look 
at some longstanding open questions that we are now able to address. 

Anniversaries present an opportunity to look both backwards and forward. An attempt is made here to capture the state of the field at this moment in time. Past achievements have been impressive indeed and are worth celebrating. The present is very very  exciting. Indeed there is a strong case to be made that the golden age of quasar research has dawned and parsec-scale observations are playing a central role. 

\section{Enablers}\label{s:enablers}
The technical side of the story of parsec-scale observations of quasars is largely a story of radio interferometry 
particularly the technique of Very Long Baseline Interferometry (VLBI). Let us be clear, VLBI is no more 
the sole source of our knowledge of parsec-scale physics of quasars than the goal-scorer is for the 
victory of a soccer team. However, it is difficult to see how many of the key parsec-scale properties summarized 
in Section~\ref{s:properties} could have been inferred without this technique. 

\subsection{The search for high resolution}\label{s:resolution} 
Given that  the radio band is at the long-wavelength end of the electromagnetic spectrum and that the resolution of all diffraction limited instruments goes as wavelength, the phrase `high-resolution radio astronomy' started life as an oxymoron. The first radio observations, Karl Jansky's measurements of cosmic radio emission had a resolution of $30^{\circ}$ \citep{1933Jansky}! And yet, radio observations are the {\it only} direct way of observing phenomena with $\sim$ milliarcsecond resolution. The great enabler here is, of course, the atmosphere, with a significant supporting role played by the fact that antennas and electronic systems behave a lot better at long wavelengths. 
Atmospheric fluctuations typically occur on timescales of a minute at radio wavelengths, four orders of magnitude slower than at optical wavelengths for example. This has made possible the development of a series of sophisticated observing and data processing techniques that make the magic of VLBI possible. 

The first improvements in radio resolution were made simply by observing at shorter radio wavelengths. For example, the first map of  the radio galaxy Cygnus\,A already had an order of magnitude better resolution than Jansky had achieved \citep{1948Reber}. This was followed by the building of larger telescopes \citep[e.g. the 76\,m Lovell telescope;][] {1957Natur.180...60L} until limits imposed by engineering and cost  were reached. Indeed, even today, fully steerable telescopes remain at the 100\,m level. Some telescopes used innovative designs to achieve higher resolution \citep[e.g. the 305\,m Arecibo telescope;][]{1964Sci...146...26G} while others were designed to take advantage of lunar occultation techniques \citep[e.g. the Ooty Radio Telescope;][]{1971ApL.....9...53S}. But the game changer was the development of interferometry, first connected element interferometers and then VLBI. 

Thanks to a range of ingenious ideas both in hardware and signal processing \citep[for an excellent review see][]{2001ARA&A..39..457K} radio interferometry, where a large effective aperture is synthesized from a modest number of (ideally) well distributed telescopes, quickly established itself as the method providing the highest resolution available in astronomy. The first parsec scale measurements made with just two antennas \footnote{Thus very limited in the information they could provide, an interferometer acts as a kind of `spatial filter' with each baseline sensitive only to information on a particular scale. A range of baselines is required to produce a high fidelity image and is typically synthesized using an array of antennas that observe over time to take advantage of the rotation of the earth.} already showed very compact features often in association with more complex structures \citep{1967Natur.215...38B, 1967ApJ...149L.151C}. Some evidence for superluminal motion soon followed \citep[e.g.][]{1971Sci...173..225W} but the quality of data had to improve considerably before their existence was convincingly established \citep{1977Natur.268..405C}. 

By the mid-eighties, VLBI was demonstrated with baselines longer than the Earth's diameter by NASA's Tracking and Data Relay Satellite System \citep[TDRSS; e.g.][]{1990ApJ...358..350L} and space VLBI techniques were further developed by the Japanese led Variable Space Orbiting Mission \citep[VSOP;][]{2000PASJ...52..997H}. The recently launched Russian mission Radioastron \citep{2012SoSyR..46..458A, 2012SoSyR..46..466A} should provide the longest baselines ever obtained. Due to the lack of intermediate baselines between the long baselines to the spacecraft and the (relatively) short intra-Earth baselines, imaging using space-VLBI is very challenging \citep[though possible e.g. see][] {2004ApJS..155...33S}. Thus, so far the biggest contribution of space-VLBI has been the measurement of very high brightness temperatures. Brightness temperatures above the well known inverse Compton cooling limit of $10^{12}$K \citep{1969ApJ...155L..71K} are of great interest as they would confront the underlying assumptions i.e. that we are observing incoherent synchrotron radiation from a cloud of relativistic electrons. Brightness temperatures exceeding the Compton limit could be most simply explained by the presence of Doppler boosting which is known to be common in many types of quasars (see below) or if the source is not in equilibrium \citep{1992ApJ...391..453S}. Contributions from more exotic mechanisms such as non-simple geometries, coherent emission processes and relativistic proton emission remain intriguing possibilities.

China has outlined a very ambitious Space VLBI program that includes multiple orbiting antennas that could make high fidelity imaging possible. This program envisions two cm-wavelength space telescopes in 2016-2020, three mm-wavelength telescopes in 2021-2025 and four submm telescopes after 2026. These telescopes are planned to be 10 to 15\,m in diameter, and will be placed in highly elliptical orbits with an apogee of 60,000\,km. The possibility of a VLBI station on the Moon is also being considered.

\begin{figure}
\centerline{\includegraphics[width=17cm]{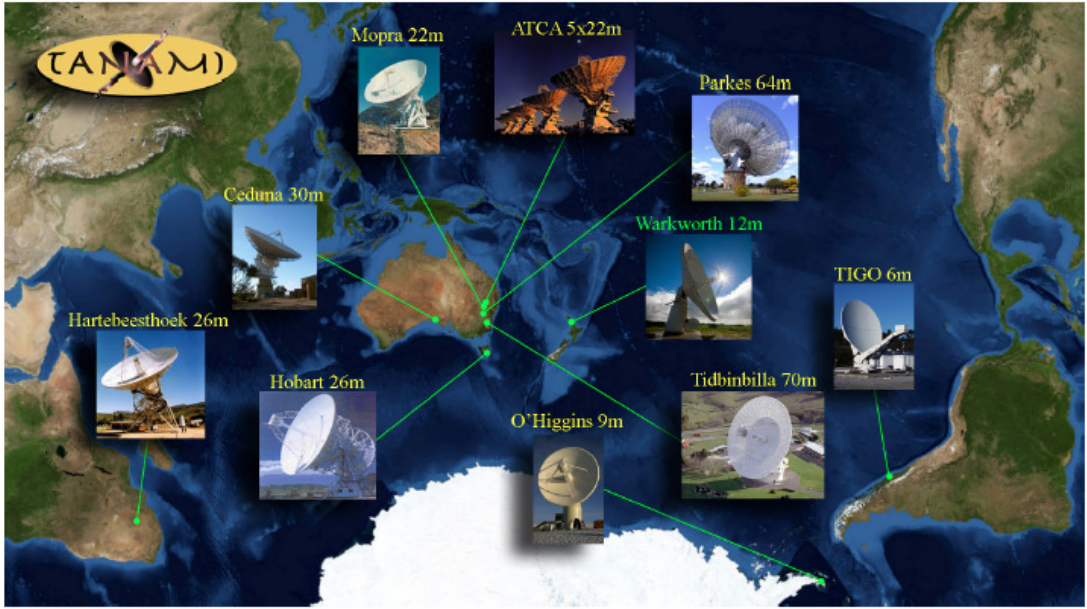}}
\caption{The TANAMI array: the telescopes of the Australian Long Baseline Array are augmented by NASA's Deep Space Network antenna in Tidbinbilla and telescopes in Antarctica, Chile, New Zealand and South Africa \citep[from][]{2013arXiv1301.5198K}. This results in the highest resolution and highest fidelity images of sources in this relatively poorly studied part of the sky. \label{f:one}}
\end{figure}

\subsection{State of the art}\label{s:stateofart} 
With any technology/technique there is a lag between its invention and its widespread use to produce robust results.
VLBI observations have been possible since the sixties and the parsec-scale information they made available immediately started impacting our understanding of quasars. However, there were several problems that made interpretation of early VLBI data tricky. Early ad-hoc arrays had to be laboriously organized and the logistics of recording and shipping equipment to and from participating telescopes meant that observations took a lot of time and effort. This obviously limited both the number of sources observed as well as the frequency and number of observations of individual sources. The number of telescopes participating tended to be small and their locations non-ideal so the resulting images were of limited quality. Few multi-epoch observations could be made with the same array and different groups using different data reduction techniques further complicated the interpretation of results. It was difficult to define source samples in a meaningful way. Thus, statistically meaningful and robust individual results had to wait for more organized VLBI arrays. There are three major such arrays used for astronomy today: the Very Long Baseline Array \citep[VLBA;][]{1994IEEEP..82..658N}, the European VLBI Network \citep[EVN;][]{1997VA.....41..303S} and the Long Baseline Array \citep[LBA;][]{1989AJ.....98....1P, 2004AJ....127.3609O}. In addition, the International VLBI Service (IVS) also organizes a worldwide network of antennas to perform regular observations of quasars for astrometric and geodetic purposes.  

The VLBA is a `bespoke' array. It was designed for VLBI with identical antennas (making calibration more reliable) optimally placed to provide the best imaging fidelity. It is also the only full time VLBI array. Thus it is the gold standard and many of the most robust results come from VLBA observations. The ongoing MOJAVE program \citep{2009AJ....137.3718L}, which monitors about 300 sources and originated in the VLBA 2cm survey \citep{1998AJ....115.1295K} in 1994, is undoubtedly the most comprehensive source of parsec-scale information on quasars available today. The EVN incorporates several extremely sensitive antennas and thus is often used to observe fainter targets. It has also been a leader in millimeter VLBI observations and eVLBI, which has the potential to eliminate the lag between VLBI observations and results, see \citet{2013ATel.4750....1F}. The Australian LBA is the only VLBI array that can observe targets south of about $- 30$ degrees declination, a part of the sky that includes many of the most interesting quasars as well as the Galactic center. The TANAMI program \citep{2010A&A...519A..45O} monitors about 85 sources in the southern third of the sky with the LBA augmented by telescopes on other continents (see Fig.~\ref{f:one}). 

To close out this section, it is worth mentioning that major new VLBI facilities are coming online in several places particularly in China, Japan and South Korea. Many of the antennas in these arrays bring significant new capabilities (e.g. the Chinese antennas are very sensitive, many of the Japanese antennas have very good high frequency performance) that will continue to improve the quality of parsec-scale information available to astronomers. 


\section{Parsec-scale properties}\label{s:properties}
Two caveats must be kept in mind while reading this section. First, though we are focused here on parsec-scale properties, it is essential to place these results in the context of multi wavelength observations to properly grasp their significance. Indeed it is the greatly enhanced availability of such broadband observations that makes this such an exciting time in quasar research. The second caveat is that such a brief review can only scratch the proverbial surface; there are numerous papers about what is covered in each of the following paragraphs. Recent books that address both caveats very well (and have comprehensive references) include \citet{2012agn..book.....B} and \citet{2012rjag.book.....B} while the well known books \citet{1997iagn.book.....P}, \citet{1999qagn.book.....K} and \citet{1999agnc.book.....K} remain extremely useful resources. 

Up to now, the word `quasar' has been used rather generally as the previous sections apply, more or less, to all active galactic nuclei. From here on we will use the word `quasar' to refer to the flat spectrum sources that are the most violently variable subset in the family of AGN and are also known as blazars. Blazars are very luminous objects and typically show high levels of radio polarization \citep[e.g.][]{1995PASP..107..803U}. Based on their optical properties these quasars/blazars are divided into two categories: BL\,Lacertae (usually just referred to as `BL\,Lac') objects and flat spectrum radio quasars (FSRQs). The former objects show weak or no broad emission lines while the latter show strong emission lines (the original formal division was based on a line equivalent width of 5\AA). With their high variability blazars form the most ``active" subset of AGN. 

Multiple lines of evidence suggest that quasars are powered by the accretion of matter onto a supermassive black hole (or, possibly, black holes) that have a mass of order $10^{6}-10^{9} M_{\odot}$. These black holes lie at the mass center of, predominantly, elliptical galaxies. The accreted matter forms a disk with a strong polar magnetic field. Dissipative processes in the accretion disk transfer matter inwards, angular momentum outwards and heat up the disk. Magnetic field lines from the inner part of the accretion disk cross the event horizon of the black hole and are wound up by its spin launching Poynting flux dominated outflows that have come to be known as `jets'  \citep{1977MNRAS.179..433B}. The inner part of the disk also 
launches a disk-wind through the so-called Blandford-Payne mechanism \citep{1982MNRAS.199..883B} which collimates these jets. 

Even parsec-scale observations are unable to directly resolve the central region of quasars, though observations at high radio frequencies are moving tantalizingly close to directly imaging the event horizon at least for the closest objects \citep{2012Sci...338..355D}. Generally, current high resolution observations show a bright, variable, compact component with a flat radio spectral index $\alpha$: $\alpha \sim 0$ where $S_{\nu} \propto \nu^{-\alpha}$. This component is identified with the center of activity in quasars and is called the `core'. As the core is the region where the jet becomes optically thick, its extent is frequency dependent and it should be kept in mind that the true center of activity is upstream \citep[e.g. see][]{2011A&A...532A..38S}. The flat radio emission of the core results from the superposition of multiple components \citep{1981ApJ...243..700K}. Cores are generally polarized at a few percent level for quasars, with BL\,Lacs cores showing somewhat higher fractional polarization than FSRQs. Radio galaxies have less than 1\% linear polarization. 

\begin{figure}
\centerline{\includegraphics[width=13.5cm]{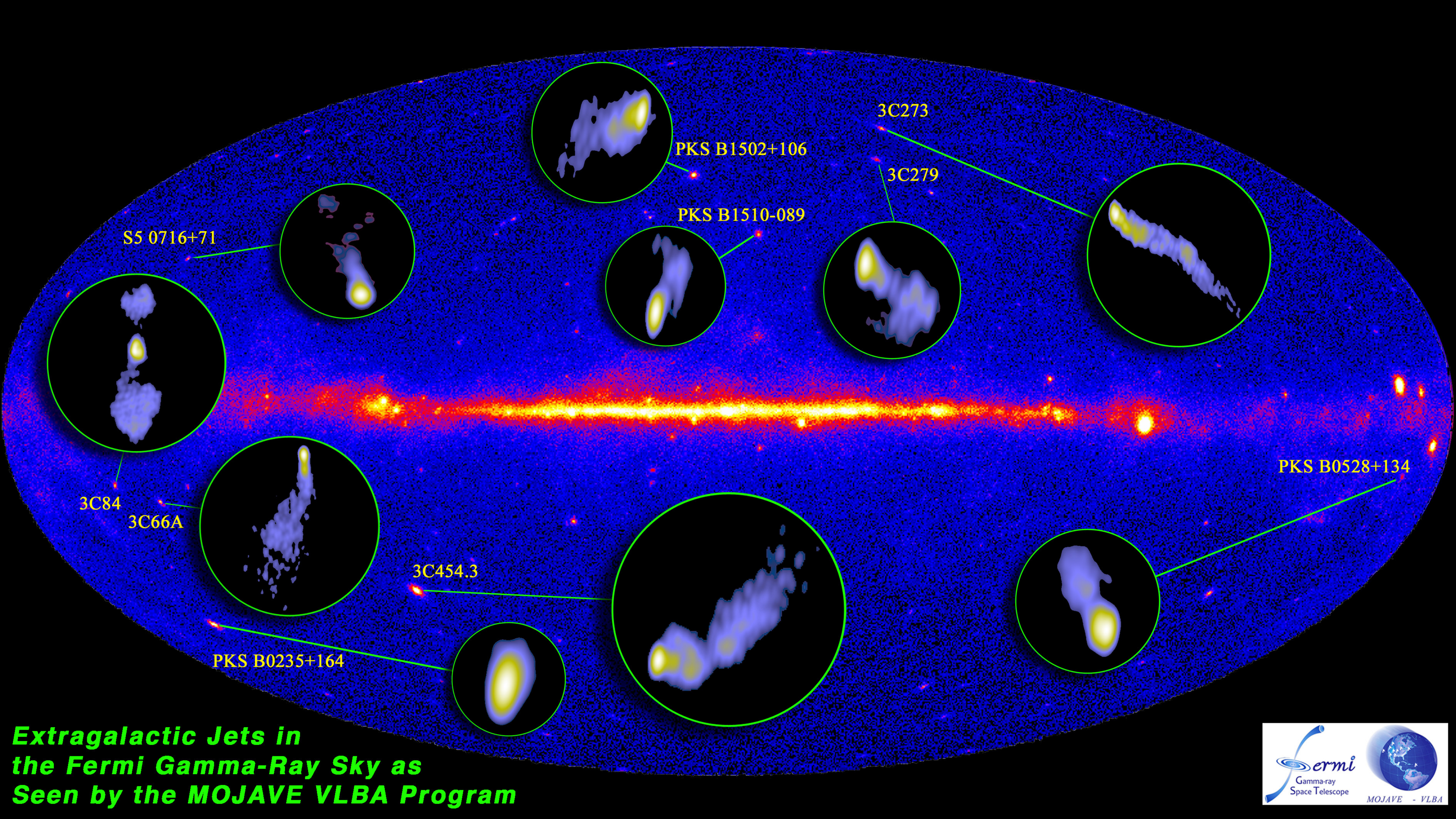}}
\caption{VLBA images of a few well known, gamma-ray loud, extragalactic sources on a background image of the gamma-ray sky. The gamma-ray image is based on the first 11 months of data from {\it Fermi}. The VLBA images are zoomed in by a factor of about 1000. Image credit: Matthias Kadler, Universit\"{a}t W\"{u}rzburg. \label{f:two}}
\end{figure}

In the canonical picture \citep{1974MNRAS.169..395B}, quasar jets are intrinsically symmetric twins, accelerating a significant fraction
of a solar mass to near light speeds on opposite sides orthogonal to the plane of the accretion disk. However, some 95\% of quasars have
an asymmetric core-jet structure (Fig.~\ref{f:two}). This is believed to result from differential Doppler boosting of intrinsically symmetric jets. If a jet is approaching an observer at a line of sight angle $\theta$ with a velocity $\beta$ it has a Doppler factor $\delta = (\Gamma(1-\beta cos\theta)^{1/2})^{-1}$. Then the approaching jet has its flux density boosted by a factor between $\delta^{2+\alpha}$ and $\delta^{3+\alpha}$ \citep{1979Natur.277..182S}. The receding jet (usually called a counter jet) is deboosted by the same factor. Not only does this make the object look like it has a jet only on one side, it also makes them appear to be far brighter than would be possible from synchrotron radiation from a stationary source without requiring unrealistically high energy densities \citep{1974ApJ...188..353J}.

Quasar jets are bright emitters across some 13 orders of magnitude in frequency, right across the electromagnetic spectrum. They are highly variable both in flux and polarization. A number of VLBI surveys have targeted large samples of quasars revealing a complex variety of morphologies and behavior \citep[e.g.][]{1981ApJ...248...61P, 1988ApJ...328..114P, 1998AJ....115.1295K, 2004ApJ...609..539K, 2004ApJS..150..187O} from which a somewhat coherent picture can be assembled. Quasar jets are sometimes straight but some degree of curvature is common. The curvature can be gentle but jets often show sharp bends sometimes exceeding $90^{\circ}$ close to the core. Many jets show multiple bends that may be a result of an intrinsically twisted (possibly helical) jet seen only at those points that are moving towards the observer and are thus Doppler brightened. The collimation of a quasar jet can stay fixed but changes are common with many examples of both narrowing and widening of jets reported in the literature. Collimation appears to be in place at the highest linear resolutions or in terms of Schwarzschild radii \citep[e.g.][]{2011A&A...530L..11M}. Typically, straight jets are fairly well aligned with their larger kiloparsec-scale counterparts, but misalignments are known. Indeed a bimodal distribution of alignments with peaks at $0$ and $90$ degrees has been claimed which would require two populations of jets, one with slightly distorted jets and a second with small Lorentz factor and large bends \citep[e.g.][]{1996A&A...310..419A}. 

The detailed structure of jets revealed by high dynamic range parsec-scale observations shows continuous emission as well as discrete structures referred to as `components' or `knots'. Jets can have many components and these components have a steeper spectral index ($\alpha \sim -0.5$) than seen for the core. New components are seen to emerge from the unresolved core region and move away from it. Different individual components in the same jet can move at different apparent speeds, though the speeds tend to be similar perhaps tracing an underlying continuous jet. Examples of both acceleration and deceleration of jet components have been seen. Different components can also move along different tracks in the same jet. Stationary components are seen in some jets but there are no convincing examples of a component moving `inwards' towards the core. Since quasar jets are aligned close to the line of sight, the radiating particles moving at relativistic speeds almost catching up with the radiation. Thus the {\it apparent} transverse motion can be greater than the speed of light. The phenomenon of `superluminal' motion is a strong confirmation of our canonical picture of quasars. Typical apparent jet speeds range from about $0$ to $15c$ though much higher speeds are seen in some sources \citep[e.g.][]{2009AJ....138.1874L}. 

Jets have a much higher degree of linear polarization than cores. Though the range is quite wide, on average BL\,Lac and FSRQ jets have higher fractional polarization than radio galaxies (up to 30\% and up to 10\% respectively). This is probably the result of greater depolarization since galaxies are seen through a thicker layer of material (the interstellar material of the host galaxy) that depolarizes the radiation through Faraday rotation \citep{1988Natur.331..147G}. 

About 5\% of sources do not show a core-jet morphology suggesting the jet axis is close to the plane of the sky or the jet has a small Doppler factor. They exhibit symmetric emission on either side of a compact core (the core can be faint and is not detected in some objects) and are referred to as Compact Symmetric Objects (CSO). CSOs are small, less than a  kpc in extent. This leads to two scenarios: either they are young objects that will eventually evolve into FRII sources or they are `frustrated' objects unable to grow due to confinement by very dense plasma. As it happens, their ages can be determined by their spectra \citep[using the break frequency distribution across their lobes e.g.][]{2008ASPC..386..290N} or kinematically \citep[by measuring the speed of advance of their hotspots e.g.][]{2004AJ....127.1977O, 2005ApJ...622..136G}. Both methods suggest they are young objects ($< 10^{4}$ years) which in turn implies they remain active for only a few hundred years. Doppler boosting does not appear to play a significant role as they are oriented far from the line of sight. This, of course, means that the counterjet and jet have comparable flux densities raising the possibility that the obscuring torus postulated by unified schemes could be detected \citep{2001ApJ...554L.147P}. Aside from a few exceptions \citep{2007ApJ...661...78G}, CSOs are not significantly polarized which limits the use of Faraday rotation techniques to study their environment. CSOs show very little flux variability and thus make very good flux density calibrators \citep{2001AJ....122.1661F}. So far there is no convincing detection of $\gamma$-ray emission from a CSO. 

Not only are jets fascinating laboratories offering insight into relativistic physics, they are also involved in the regulation of star formation and galaxy evolution via AGN feedback \citep{2007ARA&A..45..117M}. Thus they have been intensely studied and some key results have been mentioned here. It is important to note, however, that despite decades of observations and modeling, the fundamental questions on their composition, formation, collimation and dissipation are still open. It is also clear what is needed if we are to settle these opens questions: simultaneous, multiwavelength observations across the spectrum. Broadband observations in the \textsl{Fermi} era directly address several of these outstanding issues.


\section{Quasars and the International Celestial Reference Frame}\label{s:icrf}
One of the most crucial contributions that parsec scale observations of quasars have made to astronomy is the establishment of the International Celestial Reference Frame \citep[ICRF;][]{1998AJ....116..516M}. The ICRF is the realization of the International Celestial Reference System (ICRS) at radio wavelengths. Starting from 1998 January 1, the ICRF replaced the FK5 stellar catalog as the fundamental celestial reference frame for all of astronomy. The ICRF was created and is maintained by VLBI observations of compact radio sources (mostly quasars) at 8.4 and 2.3\,GHz (S and X-band respectively). It has a precision of approximately 20 microarcseconds per coordinate axis. 

To improve access to, and to control local deformations of the ICRF suitable new sources continue to be added to it particularly in the southern third of the sky where the ICRF has both a less dense distribution of sources and where the source positions are generally less well determined \citep{2006AJ....132.1944F}. Since variability in flux and morphology introduce errors in the observable quantities (group delay and delay rate), ICRF quasars are monitored to ensure their continuing astrometric suitability \citep{2000ApJS..128...17F, 2005AJ....130.2529O, 2009ITN....35....1M}. Future improvements to the ICRF include the use of very fast slewing telescopes to improve calibration, possibly moving observations to a higher frequency and using properties like scintillation as a proxy for astrometric suitability \citep{2007dpmu.book..610O}.

\section{Quasars and high energy emission}\label{s:gamma}
The EGRET instrument abroad NASA's Compton Gamma Ray Observatory satellite \citep[e.g.][]{1995MNRAS.273..583D} found that many blazars are sources of $\gamma$-ray emission \citep[e.g.][]{1992ApJ...385L...1H} thus indicating a strong link between the physics of blazars and gamma-ray emission. 
\textit{Fermi} Gamma-ray Space Telescope \citep{2009ApJ...697.1071A} observations of hundreds of blazars have 
unequivocally established the primacy of high energy emission in the energy budget of blazars.  Since late 2008, when 
it started routine observations, the LAT (Large Area Telescope) instrument on \textit{Fermi} has found blazars to be, by far, the largest category of \textsl{identified} sources in the \textsl{Fermi}/LAT First Source Catalog \citep[1FGL;][]{2010ApJS..188..405A} and in the \textsl{Fermi} LAT Second Source Catalog \citep[2FGL;][]{2012ApJS..199...31N}. It would be fair to say that one cannot really understand the physics of AGN in the absence of high energy observations.

   
Parsec scale observations are critical to broadband quasar studies for two simple reasons. VLBI monitoring provides the only direct measure of relativistic motion in AGN jets. As such, measurements of intrinsic jet parameters like speed, Doppler factor, opening and inclination angle are only possible with multi-epoch VLBI observations. These parameters are necessary inputs to models of jet behavior and to attempts to model blazar emission \citep[e.g.][]{2007ApJ...658..232C}. Further, VLBI is the only technique that can directly constrain the location and extent of the high energy emission region \citep{2012ApJ...758L..15D}. Structural changes, such as jet-component ejections, can be directly associated with $\gamma$-ray activity and flux and polarization changes at other wavelengths \citep[e.g.][]{2013hsa7.conf..152A}.
 
Before we take a quick look at what has been learned about high energy properties of blazars using parsec scale observations,  let us introduce one useful classification \citep{2010ApJ...716...30A} that is based on the location of the synchrotron emission peak, $\nu_{s}$, on a $\nu F_{\nu}$ plot. Those with $\nu_{s} < 10^{14}$ Hz are called LSP (low synchrotron peaked; these are typically radio selected), those with $10^{14} < \nu_{s} < 10^{15}$ are ISP (intermediate synchrotron peaked; these are typically X-ray selected), while those with $10^{15} < \nu_{s}$ are labeled HSP (high synchrotron peaked; these are also typically X-ray selected). Virtually all the FSRQs are LSPs. 

Blazars detected by EGRET were found to have higher radio flux and greater variability \citep{1996AJ....112.2667I}. They were also shown to be more compact and higher brightness temperatures were inferred for them \citep{1996AJ....111.2174M}.
\textit{Fermi} observations show that $\gamma$-ray flux is indeed correlated with the VLBI flux density \citep[e.g.][]{2012A&A...537A..32A} though this correlation is not present at lower radio flux levels \citep{2011ApJ...726...16L}. The $\gamma$-ray loud sources have brighter VLBI cores with higher brightness temperatures than the cores of undetected objects \citep{2009ApJ...696L..17K}. 

EGRET sources were shown to have higher apparent speeds than that of $\gamma$-ray quiet sources \citep{2001ApJS..134..181J, 2004ApJ...609..539K} though \citet{2007AJ....133.2357P} did not see this effect. This supported the idea that $\gamma-$ray loud blazars have higher Doppler factors. {\it Fermi}-LAT data show that $\gamma$-ray loud blazars do have faster jets (on average) than those not detected. Sources with significant LAT variability appear to have faster jets than those that do not vary. BL\,Lacs, which have a higher LAT detection rate than FSRQs, appear to have slower jets than FSRQs which implies that BL\,Lacs are intrinsically producing more $\gamma$-ray emission than FSRQs \citep{2009ApJ...696L..22L}. 
Higher Doppler factors for $\gamma$-ray loud blazars are also found by combining parsec scale VLBI data with single dish millimeter wavelength data \citep{2010A&A...512A..24S}. These studies also show differences in co-moving frame viewing angle distributions with a significantly narrower distribution seen in LAT-detected sources 

Parsec scale observations of EGRET sources suggested possible association of superluminal component ejection with enhanced $\gamma$-ray activity \citep{2001ApJ...556..738J}. With its essentially continuous monitoring of the whole sky (it covers the full sky once very 3 hours), LAT has tremendously expanded the possibility of studying such associations and several have been reported \citep[e.g.][]{2012JPhCS.355a2032A}. Using the VLBI core flux, instead of the emergence of a superluminal component, as a proxy for activity at radio wavelengths \citet{2010ApJ...722L...7P} found that in most cases their sample of 183 AGN were in an active radio state within 1 to 8 months of $\gamma$ ray activity. This is an intensely active field because of the salience the relative timing of such an association to the origin of high energy emission in quasars which is discussed in section ~\ref{location} below.


EGRET detections were found to have larger than average opening angles \citep{2007ApJ...671.1355T}. 
LAT-detected blazar jets appear to have larger opening angles than those not detected \citep{2009A&A...507L..33P, 2010A&A...519A..45O}. This implies two possible scenarios. Since the width of the relativistic beaming cone $\sim 1/\Gamma$, the LAT-detected jets could have smaller Lorentz factors. Or the LAT-detected jets are pointed closer to the line of sight. The former scenario is extremely unlikely as the Lorentz factor of jets with higher $\gamma$-ray flux is in fact higher \citep[e.g.][]{2009ApJ...696L..17K}. Analysis with larger samples \citep{2011ApJ...742...27L} confirm that $\gamma$-ray brighter jets have larger opening angles and it seems safe to conclude that the opening angles of these jets appear larger in projection due to smaller angles to the line of sight. 

Changes in core polarization angle in conjunction with $\gamma$-ray activity has been reported \citep[e.g.][]{2011MNRAS.413.1671F} though \citep{2011ApJ...742...27L} do not find any examples of this in their large sample.  \citep{2011ApJ...726...16L} find $\gamma$-ray loud AGN are highly polarized at the base of the jet but \citep{2011ApJ...742...27L} see no signs of this for the MOJAVE sample. \citep{2011ApJ...726...16L} find that radio galaxies detected in $\gamma$-rays have strong core polarizations. 
Core fractional polarization appears to increase during periods of $\gamma$-ray activity \citep{2012ApJ...744..177L}. The lack of agreement in some polarization properties may result from complicating factors such as high polarization variability and the presence of Faraday rotation effects. 

\citep{2011ApJ...742...27L} study a radio and $\gamma$-ray selected sample of the brightest AGN. They find evidence of a wide range of SED parameters in this population. For BL\,Lacs they find evidence for an universal SED shape and a synchrotron self-Compton origin for the high energy emission (see ~\ref{origin} for a discussion of high energy emission mechanisms). Their data also supports the idea that HSP BL\,Lac objects might have lower Doppler factors than ISP and LSP BL\,Lacs as well as FSRQs.

The VIPS survey of a large flux-limited sample \citep{2011ApJ...726...16L}, finds that BL\,Lacs detected by LAT have similar radio properties to those not detected by LAT. However,  $\gamma$-ray loud FSRQs may be intrinsically different than $\gamma$-ray quiet FSRQs with higher core brightness temperatures, higher core polarization and larger opening angles.  Since the differences they find between $\gamma$-ray loud and $\gamma$-quiet objects tend to be related to core properties, they suggest the $\gamma$-ray radiation could originate from the base of the jet. 

The above results are generally consistent with the basic picture of a Doppler boosted $\gamma$ and radio emission from a relativistic jet that goes back to the EGRET days \citep[e.g.][]{1995ApJ...440..525V}. 
Parsec scale observations in combination with {\it Fermi}-LAT data and broadband observations suggest that the viewing angle, intrinsic jet speed, the frequency at which emission peaks and the current activity state of a blazar all interact in yet to be fully understood ways to produce a $\gamma$-ray loud jet. One thing that is established beyond any doubt is the strong link between high energy blazar emission and parsec scale blazar properties.

\section{Into the golden age?}\label{s:goldenage}
Given the vast literature on quasars and the fairly consistent basic picture of their behavior one could be forgiven for considering quasar physics a mature field with major questions settled. However, many if not most of the most basic questions remain unanswered. Why do some objects emit $\gamma$-rays and why do objects with very similar properties in other wavebands do not? There are strong indications that Doppler boosting plays a major role here but there appears to be a complex dependence on several other factors (see section ~\ref{s:gamma}). Whether the radio and $\gamma$-ray emission have the same Lorentz factor is not certain and this has implications for the structure of jets. The process or processes which produce high energy radiation remain undetermined. There is the interesting related question of where gamma-ray emission originates: in the core, near the core, further along the jet are all possibilities. Indeed, where LAT has the resolution, $\gamma$-rays have even been detected from the lobes of an AGN e.g.  Centaurus A \citep{2010Sci...328..725A}. The relationships between flares in the radio and other wavebands and $\gamma$-ray flares are far from clear. Nor is the connection between changes in VLBI morphology (particularly component ejection), parsec scale polarization position angles, fractional polarization and $\gamma$-ray variability patterns. The composition of jets remains open: are they purely leptonic or do they have a significant hadronic component? As the number of blazars observed in $\gamma$-ray states increases, a rich phenomenology of flares is becoming apparent: there appear to be several different kinds of flares being produced even in the same blazar \citep[e.g][]{2013ApJ...763L..11C}. Even in situations where some reasonable models exist there are many unanswered questions e.g. the non-detection of $\gamma$-rays from the compact lobes of CSOs as predicted by \citet{2008ApJ...680..911S}.   

There are several reasons why the study of quasar physics is challenging. They are very distant objects at the limits of observational sensitivity and resolution. They are extremely anisotropic emitters so their orientation with respect to the observer has a big impact on the phenomenology. A range of  ``unified models" have been devised \citep[for a review see][]{1993ARA&A..31..473A} but this remains a complicating factor in the sense that it is yet another parameter in any model. Quasars are solitary objects so there is no possibility of obtaining information from studying members of a group such as is routinely done for stars in a cluster for example. They are fully ionized so spectral lines are hard to come by. \citet{1998MNRAS.299..433F} postulated the existence of a parameter that determines the physical and radiative properties of all types of blazars thus providing a simplifying framework to understand many blazar properties. However, it remains to be seen if this ``blazar sequence" is able to play the role that the H-R diagram has played in ordering our knowledge of stellar physics; indeed its very existence is in question, see \citet{2012MNRAS.420.2899}.

However, the biggest reason why there are so many unanswered questions about quasars can be summed up in one word: variability. Quasars are extravagantly variable. They vary on a very broad range of timescales and they vary at every wavelength they are observed at. In general the variations do not show obvious periodicity, indeed they may be stochastic. Variations at different wavelengths may be related to each other but often in complex and poorly understood ways. Thus, it has been clear for some time that to properly address basic questions, observations of quasars need to satisfy three challenging conditions. They need to be (1) broadband, (2) simultaneous and (3) well-sampled in time.

As it happens, the availability of $\gamma$-ray data from \textit{Fermi} coincides with the opening up of the TeV band as MAGIC, HESS and VERITAS have commenced operations. We have a suite of operational, space-based, X-ray telescopes: \textit{Swift}, Chandra, INTEGRAL, and now NuSTAR. And we have a range of optical/NIR and radio facilities to cover the low energy end of the spectrum. For the first time ever, simultaneous, broadband observations are possible and it is for this reason that we may be at the start of a golden age in the study of quasars. 

The VLBI community has enthusiastically entered this era, participating in intensive, high cadence `campaigns' on individual objects \citep[e.g.][]{2010ApJ...719.1433A} and organizing parsec scale monitoring of large, well-defined samples of blazars and other AGN. Variability gives rise to biases in blazar samples. Variability in the fluxes  in any given wavelength and the non-correlated changes in different wavebands which translates to variable SEDs are the culprits.  In this context, the ability of {\it Fermi}-LAT to observe the full sky every three hours (essentially continuous monitoring) is vital as it makes it possible to produce well-defined samples simplifying the analysis and interpretation of observations. 

Four complementary parsec scale monitoring programs are currently in operation. Three of them observe with the VLBA and are thus able to carry out full polarimetry observations of sources in the northern hemisphere and down to about $-30^{\circ}$ declination. As the successor to the VLBA 2\,cm survey, the MOJAVE program \citep{2009AJ....137.3718L} has many of the longest time baseline light curves available at parsec scales. It monitors about 300 sources at 15\,GHz and is well suited for statistical studies especially of kinematic properties. A larger (1100 sources) population of fainter (in radio) sources is observed at 5 and 15\,GHz by the VIPS survey \citep{2007ApJ...671.1355T}.  The blazar group at Boston University makes monthly observations of 34 blazars at 43\,GHz in conjunction with an optical polarization monitoring program to, among other things, pin down the location of high energy emission sites. 
The TANAMI program \citep{2010A&A...519A..45O} is the only dual-frequency parsec scale monitoring program measuring spectral index of jet features and their evolution. Using telescopes of the Australian LBA with telescopes in Antactica, Chile, South Africa and New Zealand, it monitors about 85 sources located in the southern third of the sky, the only parsec scale program to cover that part of the sky.


\subsection{Current problems and opportunities}\label{s:opportunities}
Let us look at just a couple of outstanding questions and how the current era in quasar research has the potential to address them. Note that parsec scale observations are essential to exploit each of these opportunities:

\subsubsection{The origin of high energy emission}\label{origin}
The spectral energy ($\nu f_{\nu}$) distributions (SED) of blazars have two broad maxima.  
Synchrotron emission from relativistic electrons in the jet produces the low energy peak in the radio to IR/optical/X-ray band \citep{1979ApJ...232...34B}. The origin of the high energy maxima (peaking anywhere from the MeV to the TeV band) is an open question. It could arise from inverse-Compton upscattering of synchrotron photons by the electrons which emitted them \citep[Synchrotron Self Compton model;][] {1981ApJ...243..700K} or from the inverse-Compton scattering of photons external to the jet by the relativistic electrons within the jet \citep[External Compton model;][]{1994ApJ...421..153S}. Several flavors of this EC model exist depending on where the reservoir of seed photons originate e.g. the BLR or the cosmic microwave background radiation. Yet another class of models suggest this second component arises from hadronic processes involving high-energy protons which produce neutral and charged pions that decay into $\gamma$-ray photons and neutrinos \citep[e.g.][]{2009APh....31..138B}. Quasi-simultaneous observations across the spectrum are essential to distinguish between these models.  None of the information in this paragraph is new or recent, what is special now is that for the first time in the 50 odd years of AGN research a suite of space and ground based observatories exist that are capable of providing quasi-simultaneous monitoring across the electromagnetic spectrum. And parsec scale VLBI observations are the only means of obtaining critical parameters such as jet speeds and core fluxes that are needed for properly modeling the blazar SEDs. 

In addition to providing critical parameters for SED modeling, VLBI observations are providing an additional challenge to the most popular class of models for fitting broadband SEDs, the one zone leptonic models mentioned above. These have already been found to be inadequate to fit some well sampled SEDs e.g. \citet{2012ApJ...760...69N}. Dual-frequency VLBI observations of Centaurus~A \citep{2011A&A...530L..11M} suggest multiple possible candidate sites for high energy emission. A large variety of models are under consideration but the big problem is that they all add free parameters which, of course, dramatically reduces their efficacy since the parameters can always be tweaked to fit the data. This represents a major challenge going forward. 


\begin{figure}
\centerline{\includegraphics[width=8cm]{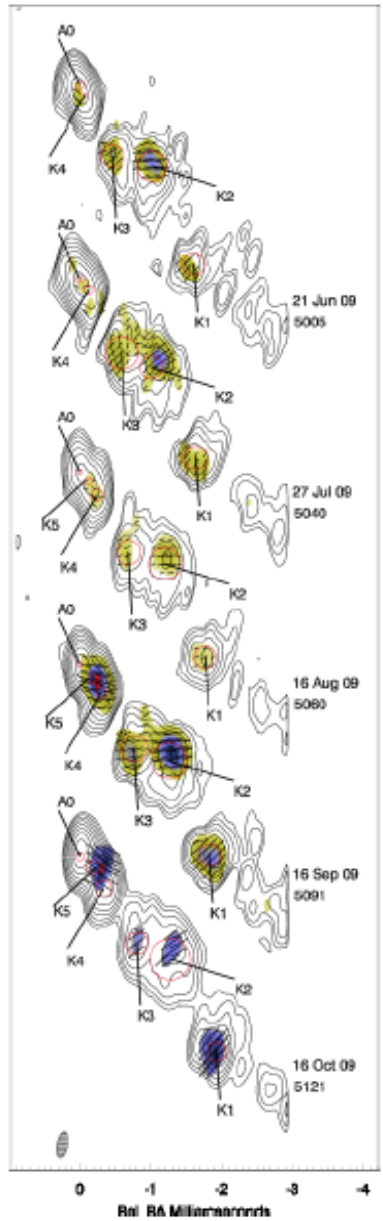}}
\caption{VLBA images at 43\,GHz of the parsec-scale jet of 3C\,273 during the early stages of the series of $\gamma$-ray flares that started around June 20, 2009 (RJD-5003). Two new superluminal knots, K4 and K5, appeared over 2 months. The linear polarization (false color) in the jet reached maximum on RJD=5091, which coincided with the highest peak in the $\gamma$-ray light curve. Figure and caption from \citet{2012arXiv1204.6707M}. \label{f:three}}
\end{figure}

\subsubsection{Location and extent of high energy emitting region}\label{location}
Identifying the location and extent of the region where high energy blazar emission originates is a fundamental question the needs to be answered if we are to understand the nature of this emission. This was already a contentious topic among theoreticians before Fermi (aided and abetted by the sparse data provided by EGRET) but has continued to generate a lot of debate since. The $\gamma$-ray emission could originate from the broad line emission region \citep[BLR;][]{2010MNRAS.405L..94T} which, at sub-parsec scales, cannot be directly resolved even by VLBI. Alternatively, the high energy emission could originate from the base of the VLBI jet seen at millimeter wavelengths or even further along the jet \citep[e.g.][]{2009ApJ...704...38S}. 

The very rapid $\gamma$-ray variability seen in several blazars \citep[variability timescales as small as $\sim 5$ mins have been seen, see][]{2007ApJ...664L..71A} provide strong constraints on the emission region size. If the emission region completely fills the cross-section of a conical jet, light travel time arguments\footnote{For a flux doubling time $t$, the size of the emitting region $R$ is: $R \leq \delta c t/(1+z)$ where $\delta$ is the Doppler factor, $c$ and $z$ have their usual meanings.} place it within a tenth of a parsec of the central engine. Since the typical BLR region is about a tenth of a parsec away this would place the $\gamma$-ray emitting region within the BLR providing a ready source of seed photons \citep[the BLR or the dust torus;][]{2011AdSpR..48..998S}
 that produce $\gamma$-ray emission by inverse Compton scattering off the relativistic jet electrons. One problem is presented by $\gamma\gamma$ attenuation that would trap high energy emission within the BLR but \citet{2012ApJ...755..147D} and \citet{2012PhRvD..86h5036T} have suggested ways this can be obviated.  

Tentative results based on EGRET data \citep{1995A&A...297L..13V} suggested that  flares at high radio frequency (i.e. the optically thin part of the radio spectrum) precede those in $\gamma$-rays which implies that the $\gamma$-ray emission originate from inverse Compton scattering of seed photons from {\it within} the jet (rather than some external source) in shocks that have boosted electron energies and magnetic field strengths. Post Fermi, much more robust data from a millimeter VLBI monitoring campaign of 34 blazars shows about two-thirds of $\gamma$-ray flares coincide with either a radio flare or a VLBI component (Fig~\ref{f:three}) or both \citep{2012arXiv1204.6707M} rather more firmly establishing this long suspected link between opposite ends of the electromagnetic spectrum. Using these VLBI measurements (specifically the opening angle, core size and linear polarization) with optical polarization and $\gamma$-ray data \citep{2011ApJ...726L..13A} show the high energy emission in OJ\,287 must have originated at least 14 parsecs away from the base of the jet. Further, optical polarization observations that show polarization position angle swings over several days \citep{2010ApJ...710L.126M} suggest the high energy emission could originate anywhere from the central engine to $~ 20$ parsecs away. So there is considerable observational evidence that the $\gamma$-ray emitting region originates well beyond the BLR. However, the statistics of the relative timing of $\gamma$-ray and radio flares/component ejections is not established beyond doubt and the fraction of sources where such correlation between radio and $\gamma$-ray emission is seen also remains to be established. 

Of course, if the high energy radiation does originate so far from the central engine there is the serious problem of explaining how the observed rapid $\gamma$-ray variability can occur. \citet{2012arXiv1204.6707M} suggest a range of possible ideas which underline the fact that this is a very active area of theoretical and observational research. 
However this argument develops, it is clear that high resolution VLBI observations (particularly at the higher radio frequencies) will be central to resolving it. VLBI observations are directly imaging these shocked regions that could be the birthplace of high energy emission. They are providing direct constraints on size, flux variability, magnetic field changes and kinematics of these regions enabling the quantitative modeling of models for high energy emission.



\section{Conclusions}\label{s:conclusion}
The age of simultaneous, broadband observations of quasars is here - it is possible to monitor quasar emission over 13 orders of magnitude in frequency. Like an orchestra beginning a performance, the various strands of observational and theoretical research are coming together to address these awe-inspiring objects. Parsec scale observations have a special, indispensable, role to play as indeed they have had in the past half century. There have been so many truly impressive results and breakthroughs in quasar research that have had far reaching effects on all astrophysics (the ubiquity of jets for example). However, it seems safe to predict that the next fifty years will be equally exciting as we use our multi-wavelength armory to tackle the remaining mysteries of quasars. 



\section*{Acknowledgements}

The author gratefully acknowledges his colleagues in the TANAMI 
and {\it Fermi/LAT} programs from whom he continues to learn. The 
author sincerely apologizes for the vast range of wonderful work that 
he was unable to touch upon in such a brief review. 
This research was funded in part by NASA through Fermi Guest
Investigator grant NNH10ZDA001N (proposal number 41213).
This research was supported by an appointment to the NASA
Postdoctoral Program at the Goddard Space Flight Center, administered
by Oak Ridge Associated Universities through a contract with NASA.
This research has made use of NASA's Astrophysics Data System.
This research has made use of the NASA/IPAC Extragalactic Database
(NED) which is operated by the Jet Propulsion Laboratory, California
Institute of Technology, under contract with the National Aeronautics
and Space Administration.
This research has made use of the SIMBAD database (operated at CDS,
Strasbourg, France).


\appendix


\label{lastpage}
\end{document}